\documentclass[preprint,superscriptaddress,double-spaced,floatfix,longbibliography]{revtex4-1}
\usepackage{graphicx,amsmath,bbm,bm,array,subfigure}
\usepackage{appendix}
\usepackage{lipsum}
\usepackage{dsfont}
\usepackage{calrsfs}
\usepackage{lineno}   
\usepackage{bbold}
\usepackage[linktocpage=true,colorlinks=true,linkcolor=blue,citecolor=blue,urlcolor=blue]{hyperref}
\usepackage{MnSymbol}
\def\nn{\nonumber\\}
\newcommand\Snn{\mathcal{S}_{nn}(\vec{k},\omega)}

\begin{document}
\title {Correlation of density fluctuation in a magnetized QCD matter near the critical end point}
\author{Mahfuzur Rahaman}
\email{mahfuzurrahaman01@gmail.com }
\affiliation{Department of Physics, Darjeeling Government College, Darjeeling- 734101, India}
\author{Md Hasanujjaman}
\email{jaman.mdh@gmail.com}
\affiliation{Department of Physics, Darjeeling Government College, Darjeeling- 734101, India}
\author{Golam Sarwar}
\email{golamsarwar1990@gmail.com }
\affiliation{Department of Physics,University of Calcutta, 92, A.P.C. Road, Kolkata-700009, India}

 \author{Abhijit Bhattacharyya}
\email{abhattacharyyacu@gmail.com}
\affiliation{Department of Physics,University of Calcutta, 92, A.P.C. Road, Kolkata-700009, India}
\author{Jan-e Alam}
\email{jane@vecc.gov.in}
\affiliation{Variable Energy Cyclotron Centre, 1/AF Bidhan Nagar, Kolkata- 700064, India}
\affiliation{Homi Bhabha National Institute, Training School Complex, Mumbai - 400085, India}

\def\zbf#1{{\bf {#1}}}
\def\bfm#1{\mbox{\boldmath $#1$}}
\def\hf{\frac{1}{2}}
\def\sl{\hspace{-0.15cm}/}
\def\omit#1{_{\!\rlap{$\scriptscriptstyle \backslash$}
{\scriptscriptstyle #1}}}
\def\vec#1{\mathchoice
{\mbox{\boldmath $#1$}}
{\mbox{\boldmath $#1$}}
{\mbox{\boldmath $\scriptstyle #1$}}
{\mbox{\boldmath $\scriptscriptstyle #1$}}
}
\def \beq{\begin{equation}}
\def \eeq{\end{equation}}
\def \beqa{\begin{eqnarray}}
\def \eeqa{\end{eqnarray}}
\def \pd{\partial}
\def \nn{\nonumber}
\begin{abstract}
The dynamical correlation of density fluctuation in quark gluon plasma with a critical end point has been investigated within the scope of the  M\"uller-Israel-Stewart theory in the presence of static ultra-high external magnetic field. The dynamic structure factor of the density fluctuation exhibits three Lorentzian peaks in absence of external magnetic field- a central Rayleigh peak and two Brillouin peaks situated symmetrically on the opposite sides of the Rayleigh peak. The spectral structure displays five peaks in presence of the magnetic field due to the coupling of the magnetic field with the hydrodynamic fields in second-order hydrodynamics. The emergence of the extra peaks is due to the asymmetry in the pressure gradient caused by the external magnetic field in the system. Interestingly, it is observed that near the critical end point, all the Brillouin peaks disappear irrespective of the presence or absence of the external magnetic field.
\end{abstract}

\maketitle
\section{Introduction}
One of the  primary objectives of  heavy-ion collision (HIC)
experiments at  Relativistic Heavy Ion Collider (RHIC)
and at Large Hadron Collider (LHC) is to create and characterize a deconfined  state of
thermal quarks and gluons, called quark gluon plasma (QGP). It is
 expected that after a proper time  $\tau_0 \sim  1 fm/c$ of the collision, the
 system attains a state of local thermal equilibrium.  The evolution of the QGP in spacetime
 can be studied by using relativistic hydrodynamics~\cite{Chaudhuri:2012yt,Heinz:2013th,Gale:2013da,Romatschke:2017ejr}, 
which is a  low frequency or long wavelength 
effective theory of many body interacting systems.  
In the non-central collisions of nuclei at RHIC and LHC energies large electric and magnetic (EM) fields will
be  produced due to the electric
current generated by the accelerated motion of charged spectators, \emph{{\it{i.e.}}} protons. In non-central collisions of heavy nuclei (Au+Au or Pb+Pb)  
at RHIC and LHC energies, the transient magnetic field ($B$) can be as high as 
~$\sim 10^{17}\mbox{-}10^{18}$ Gauss)~\cite{Bzdak:2011yy,Deng:2012pc,Tuchin:2013ie,Roy:2015coa,Li:2016tel}. 
Therefore, in such a situation, it is imperative to consider the effect of time varying magnetic and 
electric fields on the characterization of the QGP. 
The survival time of the fields crucially depends on the electrical conductivity of the  QGP~\cite{Gupta:2003zh,Aarts:2014nba,Amato:2013naa}. 
We theoretically evaluate the dynamic structure factor~\cite{Stanley_book} of the system
in the present work in an idealistic scenario of non-expanding QGP with critical point under a static magnetic field. 
The inclusion of magnetic field in relativistic hydrodynamics is studied within the ambit of 
relativistic magnetohydrodynamics~(MHD) which is a self-consistent macroscopic framework
that deals with the evolution of mutually interacting charged fluid along with the EM fields. 
Recently, several authors have studied the effect of the EM fields on QGP fluid in the context of
special relativistic systems~\cite{Huang:2009ue,Huang:2015oca,Greif:2017irh,Roy:2017yvg,Gursoy:2014aka,Huang:2015oca,Inghirami:2016iru,Huang:2011dc}.

In the relativistic viscous hydrodynamics and relativistic magnetohydrodynamics, the transport coefficients, such as the shear viscosity, bulk viscosity, thermal conductivity, etc. are taken as inputs
which can be estimated from an underlying microscopic theory~\cite{Arnold:2000dr,Arnold:2003zc,Li:2017tgi, Cao:2018ews}. A straightforward extension of the non-relativistic viscous fluid dynamics
(Navier-Stokes equation) to relativistic regime (without magnetic field)~\cite{Eckart:1940te,LL} leads to
 acausal and unstable solutions~\cite{Hiscock:1983zz,Hiscock:1985zz,Hiscock:1987zz}.
These issues were addressed and resolved by M\"uller, and Israel and Stewart~(MIS)~\cite{Israel:1976tn,Israel:1979wp}
who developed a causal and stable second-order relativistic hydrodynamics. 
Here the order of the theory is dictated by different orders of gradients in the expansion  
hydrodynamic quantities $\emph{{\it{i.e.}}}$, energy-momentum tensor (EMT).  

The dynamic density correlator or the structure factor, $\Snn$, has been studied extensively in condensed matter physics by using
linear response theory~\cite{Stanley_book} within the ambit of fluid dynamics. 
The correlation of density fluctuations can be examined through the dynamic spectral structure $\Snn$ in 
wave vector ($k$), frequency ($\omega$) space. 
Ordinarily, for a fluid modelled by relativistic hydrodynamics, the $\Snn$ consists of three distinct peaks, two of them
are Brillouin ($B$) peaks, created due to pressure fluctuation at constant entropy and the other is the Rayleigh (R) peak,
originates due to thermal or entropy fluctuation at constant pressure
~\cite{Minami:2009hn,Hasanujjaman:2020zmn}.

The spectral structure has been extensively investigated experimentally in condensed matter physics to determine the speed of sound 
by using the scattering of photons and neutrons. It is observed that the position of 
the B-Peaks in $\Snn$ depends on the speed of sound and 
the width of both B and R peaks enable us to evaluate various transport coefficients (shear and bulk viscosities, thermal conductivity),
and thermodynamic response functions (specific heats). For QCD (quantum chromodynamics) matter, no such external probes are avaiable
making the direct detection of the peaks  extremely difficult, if not impossible.  
However, such investigation may shed light on the speed of sound waves, and consequently on the equation of state (EoS) 
of the QCD matter.
In the previous studies ~\cite{Minami:2009hn,Hasanujjaman:2020zmn}, the relativistic hydrodynamics is used to find out 
the correlation in density fluctuation in QCD matter without including magnetic field. 
In the present work, we have considered the relativistic MHD to study the dynamical density correlator in a baryon-rich fluid.  
We derive $\Snn$, which allows us to determine the speed of the perturbation propagating as sound waves through the QCD medium immersed in a constant magnetic field.

The manuscript is organized as follows:~In Sec.~\ref{sec:sec2}, we discuss briefly the form of energy-momentum
tensor of fluid in the presence of EM fields.  
In the next section {\it{i.e.}} in Sec.~\ref{sec:sec3}, we find the linearized equations of
MIS hydrodynamics and derive the density-density correlation function. 
In Sec~\ref{sec:sec4}, we discuss  results that stem from our analysis, and the summary and conclusion 
is presented in Sec.~\ref{sec:conclusion}.

Throughout the article, we use the natural units, $c=k_{B}=\epsilon_{0}=\mu_{0}=1$ and the metric tensor
  is $g^{\mu\nu}=$diag$\left(+1,-1,-1,-1\right)$. The time-like 
fluid four velocity $u^{\mu}=\frac{1}{\sqrt{1-v^2}}\,(1,\vec{v})$ satisfies the relation,
$u_{\mu}u^{\mu}=1$. Also, we use the following decomposition  $\partial_{\mu}\equiv u_{\mu}u_{\nu}\partial^{\nu}+(g_{\mu\nu}-u_{\mu}u_{\nu})\partial^{\nu}=u_{\mu}D+\nabla_{\mu}$. 
The fourth-rank projection tensor is defined as
$\Delta^{\mu\nu}_{\alpha\beta}=\frac{1}{2}\left(\Delta^{\mu}_{\alpha}\Delta^{\nu}_{\beta}+\Delta^{\mu}_{\beta}\Delta^{\nu}_{\alpha}\right)-\frac{1}{3}\Delta^{\mu\nu}\Delta_{\alpha\beta}$, where $\Delta^{\mu\nu}\equiv g^{\mu\nu}-u^\mu u^\nu$ is the projection operator,
such that $\Delta^{\mu\nu}u_\nu=0$. 

\section{\label{sec:sec2}RELATIVISTIC MAGNETOHYDRODYNAMICS} 
In this section, first we discuss the equation of motion of
electromagnetic (EM) field and magnetohydrodynamics. 
	\subsection{Equation of motion of the EM field}
	We start by discussing  the relativistically
	covariant formulation of classical electrodynamics. The second rank
	anti-symmetric EM field tensor $F^{\mu\nu}$, can be defined as,   
	in terms of electric field four-vector $E^{\mu}(=F^{\mu\nu}u_\nu)$, magnetic field four-vectors~ 
$B^{\mu}=(\frac{1}{2}\epsilon^{\mu\nu\alpha\beta}u_\nu F_{\alpha\beta}=\tilde{F}^{\mu\nu}u_\nu)$, and the fluid
	four-velocity $u^{\mu}$~\cite{Lichnerowicz,Anile,Kip},
	\begin{equation}
		\label{eq:emtensor}
		F^{\mu\nu}=E^{\mu}u^{\nu}-E^{\nu}u^{\mu}+\epsilon^{\mu\nu\alpha\beta}u_{\alpha}B_{\beta},
	\end{equation}
and its  dual counter part is represented by,
	\begin{equation}
		\label{eq:dual}
		\tilde{F}^{\mu\nu}=B^{\mu}u^{\nu}-B^{\nu}u^{\mu}-\epsilon^{\mu\nu\alpha\beta}u_{\alpha}E_{\beta}. 
	\end{equation}
where, $\epsilon^{\mu\nu\alpha\beta}$ is the Levi-Civita tensor. It is quite straight forward to see by using 
the anti-symmetric property of $F^{\mu\nu}$ that both $E^{\mu}$ and $B^{\mu}$
	are orthogonal to $u^{\mu}$ {\it{i.e.}}, $E^{\mu}u_{\mu}=B^{\mu}u_{\mu}=0$. It may be noted that in the
	rest frame $u^{\mu}=(1,\bf 0)$, $E^{\mu}=(0,\bf E)$, and $B^{\mu}=(0,\bf B)$, where $\bf E$ and $\bf B$
	correspond to the electric and the magnetic three vectors fields with $\mathrm{E}^{i}=F^{i 0}$ and
	$\mathrm{B}^{i}=-\frac{1}{2} \epsilon^{i j k} F_{j k}$. 
	The indices $i, j, k$ run over $1, \,2,\, 3$. The  $E^\mu$ and $B^\mu$ can be
	interpreted as the electric and magnetic fields respectively measured
	in a frame in which the fluid moves with  velocity $u^\mu$.
	
Using the  EM field tensor and its dual, we can express the  Maxwell's equations in a covariant way as,
	\begin{eqnarray}
		\label{eq:maxwelleqn1}
		\partial_{\mu}F^{\mu\nu} &=& J^{\nu}, \\
		\label{eq:maxwelleqn2}
		\partial_{\mu}\tilde{F}^{\mu\nu} &=& 0,
	\end{eqnarray}
	where, $J^{\nu}$ represents   the electric charge four-current and it acts as the source of EM field.
 The electric charge four-current ($J^\mu$) can be decomposed as follows: 
	\begin{equation}
		\label{eq:chargecurrent}
		J^{\mu}=j^{\mu}+d^{\mu},
	\end{equation}
	where, $j^{\mu}$ is the conduction current and
	$d^{\mu}=\Delta^{\mu}_{\nu}J^{\nu}$, is the charge diffusion current,
	$n_{q}=u_{\mu}J^{\mu} $ the proper net charge density.
	If we consider a linear relation between $j^{\mu}$ and $E^{\mu}$~(Ohm's law) then we can write
	$ j^{\mu}=\sigma^{\mu\nu}E_{\nu}$, where $\sigma^{\mu\nu}$ is the second rank conductivity tensor.
	The construction of $j^{\mu}$ follows $u_{\mu}j^{\mu}=0$. It implies that the conduction current
	exists even in the absence of any  net charge.
	The solutions of Eqs.~\eqref{eq:maxwelleqn1} and ~\eqref{eq:maxwelleqn2} for given electric charge four current
	$J^{\mu}$ in Eq.~\eqref{eq:chargecurrent} completely determine the evolution of 
electromagnetic field.
	It also acts as a coupling between the fluid and  EM fields as it contains the fluid
	information {\it{e.g.}} fluid conductivity $\sigma^{\mu\nu}$, net charge density $n_{q}$ etc. 
We consider a system consists of quarks, anti-quarks and gluons. All the relevant thermodynamic quantities and the EoS
used in this work have been calculated by taking these degrees of freedom into account (we do not repeat the discussions on EoS here
but refer to Ref.~\cite{Hasanujjaman:2020zex} for details).
Therefore, the net (quark-antiquark) charge density is equivalent to net (quark-antiquark) number density 
and the relation $n_{q}=q n_{f}$ holds, where $n_{f}$ corresponds to net number density.

	
	At first, we assume  that the fluid   does not possess any  polarization or magnetization. In that case  the EM field stress-energy tensor (EMT) can be written as,
	\begin{equation}\label{Temmunu}
		T^{\mu\nu}_{EM}=-F^{\mu\lambda}F^{\nu}_{\lambda}+\frac{1}{4}g^{\mu\nu}F^{\alpha\beta}F_{\alpha\beta}.
	\end{equation} 
If we take the partial derivative of the field stress-energy tensor, we obtain the equation of motions,
	\begin{equation}\label{partialTmunu}
		\partial_{\mu}T^{\mu\nu}_{EM}=-F^{\nu\lambda}J_{\lambda}.
	\end{equation}
	In the above equation the current density due to external source is ignored.
	In presence of an external source ($J^{\mu}_{ext}$),
	the total current is given by,
	\begin{equation}\label{totalchargedensity}
		J^{\mu}=J^{\mu}_f +J^{\mu}_{ext}.
	\end{equation}
	In such case, the external current   acts as a source term in   energy-momentum conservation
	equation.

	In this article, we consider an ideal magneto-hydrodynamic limit which resembles  very large magnetic Reynolds number
	$R_{m}=Lv\sigma \mu >>1$, where $L$ is the characteristic macroscopic 
length or time scale, $v$ is the characteristic velocity 
        of the flow, $\sigma$ is the isotropic electrical conductivity ($\sigma^{\mu\nu}=\sigma g^{\mu\nu}$),
and $\mu$ is the magnetic permeability of the QGP. 
It is evident that $R_m$ increases with $\sigma$. From the induced current density equation, $J^{\mu}_{ind}=j^{\mu}=\sigma E^{\mu}$, 
it is clear that when $\sigma$ attains a very large value ($\rightarrow\infty$) 
then the electric field will approach zero ($E^\mu\rightarrow 0$) to keep 
$J^{\mu}_{ind}$ but finite.
This simplifies the EM tensor $F^{\mu\nu}$ to the following form,
	\begin{equation}
		\label{eq:modemtensor}
		F^{\mu\nu}\rightarrow B^{\mu\nu}=\epsilon^{\mu\nu\alpha\beta}u_{\alpha}B_{\beta}.
	\end{equation}
	Using  Eqs.~\eqref{totalchargedensity} and~\eqref{eq:modemtensor} in
	the Maxwell's equations Eq.~\eqref{eq:maxwelleqn1} we obtain,
	\begin{equation}
		\label{eq:EoMmaxmodified}
		\epsilon^{\mu\nu\alpha\beta}\left(u_{\alpha}\partial_{\mu}B_{\beta}+B_{\beta}\partial_{\mu}u_{\alpha}\right)=J^{\nu}_{f} +J^{\nu}_{ext}.
	\end{equation}
	The EMT in absence of electric field can be obtained  
	from  Eqs.~\eqref{Temmunu} and~\eqref{eq:modemtensor} as,
	\begin{equation}\label{emtensorforB}
		T^{\mu\nu}_{EM}\rightarrow T^{\mu\nu}_B=\frac{B^2}{2} \left(u^{\mu}u^{\nu}-\Delta^{\mu\nu}-2b^{\mu}b^{\nu}\right),
	\end{equation}
	where $B^{\mu}B_{\mu}=-B^{2}$ and $b^{\mu}=\frac{B^{\mu}}{B}$ with the constraints
	$b^{\mu}u_{\mu}=0$ and $b^{\mu}b_{\mu}=-1$. Using  Eq.~\eqref{eq:modemtensor}, 
	one can show that $B^{\mu\nu}B_{\mu\nu}=2B^{2}$. We
	  can define another anti-symmetric tensor   as: $b^{\mu\nu}=-{B^{\mu\nu}}/{B}$.
	\subsection{Equation of motion of magnetohydrodynamics}
In this section, we derive the equation of motion for relativistic fluid under the influence of external EM field.
	\subsubsection{Conservation of energy and momentum of fluid and electromagnetic field}
In the absence of any magnetic field, the EMT and the particle currents
are conserved separately according to the following conservation laws,
	\begin{eqnarray}
		\partial_{\mu}N^{\mu} &=& 0, \\
		\partial_{\mu}T_f^{\mu\nu} &=& 0.
	\end{eqnarray}
In presence of external EM field 
	 the total EMT tensor (fluid+field), $T^{\mu\nu}$ is given by,
	\begin{equation}
		\label{eq:totem}
		T^{\mu\nu}=T^{\mu\nu}_f+T^{\mu\nu}_{EM},
	\end{equation}
where $T^{\mu\nu}_f $  and $ T^{\mu\nu}_{EM}$ are the contributions from the fluid and EM field 
respectively.
	The total EMT in Ref.~\cite{Anile} contains additional terms 
	which can not be unambiguously attributed to  the fluid or to the field.  But for  constant
	susceptibility and vanishing $E^{\mu}$ such contributions vanish and then Eq.~\eqref{eq:totem}
	becomes a good approximation. As the  electric
	charge is conserved, the charge current of the fluid is individually conserved too,
	\begin{equation}\label{concharge}
		\partial_{\mu}J^{\mu}_f=0.
	\end{equation}
	If we have an external charge current, then it will act as a source term of the EMT:
	\begin{equation}\label{partialtot}
		\partial_{\mu}T^{\mu\nu}=-F^{\nu\lambda}J_{ext,\lambda}.
	\end{equation}
	The conservation equation for the electromagnetic field Eq.~\eqref{partialTmunu} with external source
	can be written as,
	\begin{equation}\label{partialem}
		\partial_{\mu}T^{\mu\nu}_{EM}= -F^{\nu\lambda}\left(J_{f,\lambda}+J_{ext,\lambda}\right).
	\end{equation}
	Using Eqs~\eqref{eq:totem},~\eqref{partialtot} and ~\eqref{partialem} we get,
	\begin{equation}\label{partialfluid}
		\partial_{\mu}T^{\mu\nu}_f= F^{\nu\lambda}J_{f,\lambda}.
	\end{equation}
	Usually, the total EMT of an isolated system remains conserved but in  
	the presence of an external  charge current an appropriate 
	source term should be taken into account.  In this case, the fluid evolution depends on the
	fluid charge current density  through Eq.~\eqref{partialfluid}.
	
	It is also useful to express the conservation equations by taking
	projection along and perpendicular to fluid four velocity  in an alternative way.
	If we take the parallel projection of Eq.~\eqref{partialem} and Eq.~\eqref{partialfluid}, then we obtain,
	\begin{eqnarray} 
		\label{eq:paralEM}
		u_{\nu}\partial_{\mu}T^{\mu\nu}_{EM} &=& 0, \\
		\label{eq:paralFl}
		u_{\nu}\partial_{\mu}T^{\mu\nu}_f &=& 0.
	\end{eqnarray}
	In presence of current density given by  Eq.~\eqref{totalchargedensity},
	the perpendicular projection of Eqs.~\eqref{partialem} and ~\eqref{partialfluid} can be written as:
	\begin{eqnarray}
		\label{eq:perpEM}
		\Delta^{\alpha}_{\nu}\partial_{\mu}T^{\mu\nu}_{EM}&=& Bb^{\alpha\lambda}\left(J_{f,\lambda}+J_{ext,\lambda}\right), \\
		\label{eq:perpFL}
		\Delta^{\alpha}_{\nu}\partial_{\mu}T^{\mu\nu}_f &=& -Bb^{\alpha\lambda}J_{f,\lambda}.
	\end{eqnarray}
	It implies that  the momentum density of  fluid depends
	on  diffusion current/magnetic field,  momentum density of the field,
	external current and  fluid diffusion current. 
	
	\subsubsection{Ideal and dissipative non-resistive magnetohydrodynamics}
 	The EMT for non resistive magnetohydrodynamics is given by, 
	\begin{eqnarray}
		 T^{\mu\nu}_{EM}\equiv T^{\mu\nu}_{B}=\frac{B^2}{2} \left(u^{\mu}u^{\nu}-\Delta^{\mu\nu}-2b^{\mu}b^{\nu}\right).
	\end{eqnarray} 
	In case the ideal fluid,  the total EMT takes the form-
	\begin{equation}\label{idealmhd}
		T^{\mu\nu}_{(0)}=\left(\epsilon+\frac{B^2}{2}\right)u^{\mu}u^{\nu}-\left(P+\frac{B^2}{2}\right)\Delta^{\mu\nu}-B^2b^{\mu}b^{\nu}\,,
	\end{equation}
where, $\epsilon$  and $P$ denote energy density and the thermodynamic pressure respectively. 
For dissipative fluid with non-zero shear and bulk viscosities, and non-zero thermal conductivity, the EMT becomes-
	\begin{equation}
		{\label{eq:fullEM}}
		T^{\mu\nu} = \left(\epsilon+\frac{B^2}{2}\right)u^{\mu}u^{\nu}-\left(P+\Pi+\frac{B^2}{2}\right)\Delta^{\mu\nu}-B^{\mu}B^{\nu}+q^\mu n^\nu +q^\nu n^\mu + \pi^{\mu\nu},
	\end{equation}
where, $\Pi, q^\mu$ and $\pi^{\mu\nu}$ denote bulk pressure, heat flux and shear stress respectively, and are expressed in Eckart's frame of reference~\cite{Eckart:1940te,Israel:1979wp}. The system of equations can be closed with  the help of  constitutive relation of charged-current and with an Equation of State (EoS), which relates the thermodynamic
	pressure to energy and number density $P=P\,(\epsilon,\,n_{f})$.
	
	The inclusion of the polarization modifies the  
	energy-momentum tensor as follows~\cite{Huang:2009ue,Groot:1969,Israel:1978up}:
	\begin{eqnarray}
		\label{tmn00} T^{\mu\nu}_0&=&T_{\rm F0}^{\mu\nu}+T_{\rm EM}^{\mu\nu},\\
		\label{tmn00_1}
		T^{\mu\nu}_{\rm F0}&=&\epsilon\, u^\mu u^\nu-P\Delta^{\mu\nu}-\frac{1}{2}\left( 
		M^{\mu\lambda}F^{\;\;\nu}_{ \lambda}+M^{\nu\lambda}F^{\;\;\mu}_{\lambda}\right),\\
		\label{tmn00_2}
		T_{\rm EM}^{\mu\nu}&=&-F^{\mu\l}F_{\;\;\lambda}^\nu+\frac{g^{\mu\nu}}{4}F^{\rho\sigma}F_{\rho\sigma},
	\end{eqnarray}
 where, $M^{\mu\nu}$ is the polarization tensor.
	In the non-dissipative limit, the entropy is also conserved along with  the charge.
The charge
	and entropy currents can be expressed in  the non-dissipative hydrodynamics as,
	\begin{eqnarray}
		n_{0}^\mu&=&n\, u^\mu ,\\
		s_0^\mu &=& s\, u^\mu ,
	\end{eqnarray}
	where, $n$ and $s$ are the electric charge density and entropy
	density measured in the local rest frame.  It is convenient to decompose the tensor $F^{\mu\nu}$
	into components parallel and perpendicular to $u^\mu$ as:
	\begin{eqnarray}
		\label{fmn} F^{\mu\nu}&=&F^{\mu\lambda}u_\lambda u^\nu-F^{\nu\l}u_\lambda
		u^\mu+\Delta^\mu_{\;\;\alpha}F^{\alpha\beta}\Delta_\beta^{\;\;\nu}\nonumber\\ &\equiv&E^\mu u^\nu-E^\nu
		u^\mu+\frac{1}{2}\epsilon^{\mu\nu\alpha\beta}\left(  u_\alpha B_\beta-u_\beta B_\alpha\right),
	\end{eqnarray} 
	
	The anti-symmetric polarization tensor $M^{\mu\nu}$ represents
	the response of matter to   $F^{\mu\nu}$. It is
	given by $M^{\mu\nu}\equiv -\partial\Phi/\partial F_{\mu\nu}$, 
where $\Phi$ is the
	 thermodynamic potential function. It is also convenient to  define
	the in-medium field strength tensor $H^{\mu\nu}\equiv
	F^{\mu\nu}-M^{\mu\nu}$.  Similar to $F^{\mu\nu}$, we 
	decompose $M^{\mu\nu}$ and $H^{\mu\nu}$ as follows:
	\begin{eqnarray}
		M^{\mu\nu}&=& P^\nu u^\mu-P^\mu u^\nu+\frac{1}{2}\epsilon^{\mu\nu\alpha\beta}\left( 
		M_\beta u_\alpha-M_\alpha u_\beta\right), \\ H^{\mu\nu}&=& D^\mu u^\nu-D^\nu
		u^\mu + \frac{1}{2}\epsilon^{\mu\nu\alpha\beta}\left(  H_\beta u_\alpha-H_\alpha u_\beta\right),
	\end{eqnarray}
	with $P^\mu\equiv -M^{\mu\nu}u_\nu$, $M^\mu\equiv
	\epsilon^{\mu\nu\alpha\beta}M_{\nu\alpha}u_\beta/2$,
	$H^\mu\equiv \epsilon^{\mu\nu\alpha\beta} H_{\nu\alpha}u_\beta/2$, and $D^\mu\equiv H^{\mu\nu}u_\nu$.
	
	In the local rest frame of the fluid, the non-trivial components of these
	tensors are $(F^{10}, F^{20}, F^{30})=\bf{E}$, $(F^{32}, F^{13},
	F^{21})=\bf{B}$, $(M^{10}, M^{20}, M^{30})=-\bf{P}$, $(M^{32},
	M^{13}, M^{21})=\bf{M}$, $(H^{10}, H^{20}, H^{30})=\bf{D}$, and
	$(H^{32}, H^{13}, H^{21})=\bf{H}$, where $\bf{P}$ and $\bf{M}$ are
	the electric polarization and magnetization vector 
	respectively. In linear domain,  they are related to the
	fields $\bf{E}$ and $\bf{B}$ by the   expressions
	${\bf P}=\chi_e{\bf E}$ and ${\bf
		M}=\chi_m{\bf B}$, where $\chi_e$ and $\chi_m$ represents the electric and
	magnetic susceptibilities respectively. Here $E^\mu, B^\mu$  are
	space-like, {\it {\it{i.e.}}} $E^\mu E_\mu=-E^2$ and $B^\mu B_\mu=-B^2$ and orthogonal to $u^\mu$
{\it {\it{i.e.}}} $E^\mu u_\mu=0, B^\mu u_\mu=0$ with
$E\equiv |\bf E|$ and $B\equiv |\bf B|$.
	
There are several physical system where the electric field is much
	weaker than the magnetic field. The interior of a neutron star is one such example. 
In the following discussions,
	we  will omit the contribution  from the electric field. We   introduced  the four-vector $b^\mu\equiv
	B^\mu/B$, which is normalized as $b^\mu b_\mu=-1$  along with the anti-symmetric rank-2 tensor
	$b^{\mu\nu}\equiv\epsilon^{\mu\nu\alpha\beta} b_\alpha u_\beta$. In the absence of  electric field,
	we have,
	\begin{eqnarray}
		F^{\mu\nu}&=&-B b^{\mu\nu},\\ M^{\mu\nu}&=&-M b^{\mu\nu},\\
		H^{\mu\nu}&=&-H b^{\mu\nu},
	\end{eqnarray}
	where $M\equiv |\bf M|$ and $H\equiv |\bf H|$. In absence of
	electric fields, the matter and field contributions to
	the EMT (\ref{tmn00}) can now be written
	in terms of $b^{\mu}$ and $b^{\mu\nu}$
	as~(see, {\it {\it{e.g.}}}, Refs.~\cite{Huang:2009ue,Gedalin:1991})
	\begin{eqnarray}
		\label{tmn0} T^{\mu\nu}_{\rm F0}&=&\varepsilon u^\mu
		u^\nu-{P_\perp}\Xi^{\mu\nu}+{P_\parallel} b^\mu b^\nu,\\
		T^{\mu\nu}_{\rm EM}&=&\frac{1}{2}B^2\left(  u^\mu u^\nu-\Xi^{\mu\nu}-b^\mu
		b^\nu\right),
	\end{eqnarray}
	where $\Xi^{\mu\nu}\equiv\Delta^{\mu\nu}+b^\mu b^\nu$
	is a new projection tensor with 
	$\Xi^{\mu\nu}u_\mu=\Xi^{\mu\nu}b_\mu=0$. 
The transverse and longitudinal pressures relative to $b^\mu$ can be defined as ${P_\perp}=P-MB$ and ${P_\parallel}=P$
	relative to the vector $b^\mu$.  In the
	absence of   magnetic field, the fluid is isotropic and
	${P_\perp}={P_\parallel}=P$, where $P$ is the thermodynamic pressure defined
	in Eq.~(\ref{tmn00_2}). In the local rest frame of the fluid,	the direction of the magnetic field is chosen as the $z$-axis without loss of generality, so we have
	$b^\mu=(0,0,0,1)$. 
	Then the EMT takes the  form, $T^{\mu\nu}_{\rm F0}={\rm{diag}}(\epsilon,{P_\perp}, {P_\perp}, {P_\parallel})$.
	
	In the presence of polarization (or non-zero magnetization),  the full energy momentum tensor (EMT) for  relativistic  MHD can be expressed as,
	\begin{eqnarray}
		T^{\mu\nu} &=&(\epsilon+P-MB)u^\mu u^\nu -(P-MB+\frac{1}{2}B^2)g^{\mu\nu}+(MB-B^2)b^\mu b^\nu \nn\\
		 &&- \Pi \Delta^{\mu \nu}+q^{\mu}u^{\nu}+q^{\nu}u^{\mu}+\pi^{\mu \nu}\,.
		 \label{eq39}
	\end{eqnarray}
The  form of $q^{\mu}, \Pi, \pi^{\mu \nu}$ used in MIS hydrodynamics contains
additional coupling and relaxation coefficients arising due to inclusion of 
second order gradients~\cite{Hiscock:1983zz,Van:2007pw,Baier:2007ix}:
\beqa
\Pi&=& -\zeta\Big[\pd_{\mu}u^{\mu}+\beta_{0}D\Pi-\alpha_{0}\pd_{\mu}q^{\mu}\Big] \,,\nn\\
\pi^{\lambda \mu}&=& 2\eta \Delta^{\lambda\mu\alpha\beta}\Big[\partial_{\alpha}u_{\beta}-\beta_{2}D\pi_{\alpha\beta}-\alpha_{1}\partial_{\alpha}q_{\beta}\Big]\,, \nn\\
q^{\lambda}&=&\kappa T\Delta^{\lambda\mu}\Big [ \frac{1}{T}\partial_\mu T -Du_{\mu}-\beta_1 D{q_\mu}-\alpha_0\partial_\mu \Pi +\alpha_1\pd_{\nu}\pi ^{\nu}_{\mu} \Big]\,,
\label{eq40}
\eeqa
where, $\zeta$, $\eta $, and $\kappa$ are the coefficient of bulk viscosity, 
shear viscosity, and thermal conductivity 
respectively, and 
$\Delta^{\mu\nu\alpha\beta}=
\frac{1}{2}\big[\Delta^{\mu\alpha}\Delta^{\nu\beta}+\Delta^{\mu\beta}\Delta^{\nu\alpha}-
\frac{2}{3}\Delta^{\mu\nu}\Delta^{\alpha\beta}\big]$. Here, $\beta _0,\beta_1,\beta_2$ are relaxation coefficients, 
$\alpha_0$ and $\alpha_1$ are coupling coefficients. The relaxation times for the 
bulk pressure ($\tau_{\Pi}$), the heat flux ($\tau_q$) and the shear tensor ($\tau_{\pi}$) 
are defined as~\cite{Muronga:2006zw} 
\begin{equation}
\tau_{\Pi}=\zeta \beta_0, \,\,\,\,\tau_q=k_BT\beta_1,\,\,\,\, \tau _{\pi}=2\eta \beta_2\,.
\label{eq12}
\end{equation}
The relaxation lengths which  couple to   heat flux and bulk  pressure 
($l_{\Pi q}, l_{q\Pi} $), the heat flux and shear tensor $(l_{q\pi}, l_{\pi q})$ 
are defined as follows:
\begin{equation}
l_{\Pi q}=\zeta \alpha_0,\,\,\,\, l_{q\Pi}=k_B T \alpha _0, \,\,\,\,l_{q\pi}=k_BT\alpha_1,\,\,\,\, 
l_{\pi q}=2\eta \alpha_1 \,.
\label{eq13} 
\end{equation}  
In the ultra-relativistic limit, $\beta(=m/T)\rightarrow 0$, where $m$ is the mass
of the particle. We also have~\cite{Israel:1979wp},
\begin{eqnarray}
\alpha_0 \approx  6\beta^{-2}P^{-1},\,\,\,\, \alpha_1\approx -\frac{1}{4}P^{-1},\,\,\,\,
\beta_0 \approx 216 \beta^{-4} P^{-1},
\beta_1 \approx \frac{5}{4}P^{-1},\,\,\,\, \beta_2 \approx \frac{3}{4}P^{-1} \,.
\label{eq14}
\end{eqnarray}
	To check the consistency of the terms 
	in $T_{\rm F0}^{\mu\nu}$ 
	involving electromagnetic fields, we  use the thermodynamic
	relation,
	\begin{eqnarray}
		\label{thermo} \epsilon&=&Ts+\mu n-P\,,
	\end{eqnarray}
where, $\mu$ is the chemical potential introduced to constrain the conservation of net (baryon) number. 
	By using  the conservation equations for $n_{0}^\mu$ and $s_0^\mu$ in the ideal
	hydrodynamics, it is straight forward  to show that the hydrodynamic equation
	$u_\nu\partial_\mu T^{\mu\nu}_0=0$ along with the Maxwell equation
	Eq.~\eqref{eq:maxwelleqn1} implies,
	\begin{eqnarray}
		\label{de} D\epsilon=TDs+\mu D n-MDB,
	\end{eqnarray}
	which is in accordance  with the standard thermodynamic relation,
	\begin{eqnarray}
		\label{de2} d\epsilon&=&Tds+\mu dn-MdB.
	\end{eqnarray}
	From Eqs.~(\ref {thermo}) and ~(\ref {de2}), we obtain the Gibbs-Duhem
	relation,
	\begin{eqnarray}
		\label{gd} dP=sdT+nd\mu+MdB.
	\end{eqnarray}
	The complete set of non-dissipative hydrodynamic equations in presence of 
an external magnetic field is formed 
        by Eqs.\eqref{eq39} and \eqref{eq40} along with the following two equations:
	\begin{eqnarray}
		\label{maxwell0} \partial_\nu \left( B^\mu u^\nu-B^\nu u^\mu\right)&=&0,\\
		\label{maxwell2}\partial_\mu H^{\mu\nu}&=&n^\nu.
	\end{eqnarray}
	Contracting Eq.~(\ref{maxwell0}) with $b_\mu$, one obtains 
	the induction equation,
	\begin{eqnarray}
		\label{maxwell}\theta+D\ln B-u^\nu b^\mu\partial_\mu b_\nu=0,
	\end{eqnarray}
	where, $\theta\equiv\partial_\mu u^\mu$ is the velocity divergence and $D\equiv u^\mu\partial_\mu$ is the co-moving derivative.
	All of these facts  confirm that the form of EMT, $T_{\rm F0}^{\mu\nu}$ is consistent
	with  thermodynamic formulas for matter in the presence of
	electromagnetic fields~\cite{Huang:2009ue}.
	\section{Hydrodynamic equations in linearised form to estimate the dynamic structure factor}
	\label{sec:sec3}
	Presently, we aim to evaluate the dynamic structure factor [$\Snn$] by taking the correlation of dynamical density
	fluctuations in $\omega-k$ space. A small deviation from the equilibrium state of a thermodynamic variable can be 
accompanied by linearized form of the hydrodynamic equations. The density fluctuation can be obtained from the linearised equation 
to estimate the structure factor. Let the equilibrium state and the away from the equilibrium state of a thermodynamic quantity 
($n, \,\epsilon,\, u^{\alpha},\, q^{\alpha},\, s,\, \Pi,\, \pi^{\alpha \beta}$ etc.) are 
generically denoted by $Q_{0}$ and $Q$ respectively. Therefore, any state (away from equilibrium) can be expressed as $Q=Q_{0}+\delta Q$, where $\delta Q$ is some tiny perturbation to the equilibrium value, $Q_{0}$. We can thus express the hydrodynamic equations around the equilibrium in the linearized form as:
	\begin{subequations}
	\begin{eqnarray}
	\label{eq58a}
	0&=&\frac{\pd \delta n}{\pd t}+n_{0} \vec{\nabla}.\delta \vec{v}\,,  \\
	\label{eq58b}
	0&=& n_{0}\frac{\pd \delta s}{\pd t}+\frac{1}{T_{0}}\vec{\nabla}.\delta \vec{q}- \frac{1}{n_0 \, T_0}(B^2-M \, B)\frac{\pd \delta n}{\pd t}\,,\\
	\label{eq58c}
	0&=& (h_{0}+B^2-M \, B) \frac{\pd \delta v}{\pd t}+\nabla(\delta P+ \delta \Pi)+\frac{\pd \delta q}{\pd t}+\vec{\nabla}.\delta\vec{\pi}\,, \\
	\label{eq58d}
	0&=& \delta \Pi +\zeta\Big[\vec{\nabla}.\delta \vec{v}+\beta_{0}\frac{\pd \delta \Pi}{\pd t}-\alpha_{0}\vec{\nabla}.\delta \vec{q}\Big] \,,\\
	\label{eq58e}
	0&=& \delta \pi^{ij}-\eta\Big[\pd^{i}\delta v^{j}+\pd^{j}\delta v^{i}-\frac{2}{3}g^{ij}\vec{\nabla}.\delta \vec{v}-2\beta_{2}\frac{\pd \delta \pi^{ij}}{\pd t} -\alpha_{1}(\pd^{i}\delta q^{j}+\pd^{j}\delta q^{i}-\frac{2}{3}g^{ij}\vec{\nabla}.\delta \vec{q})\Big]\,,\\
	\label{eq58f}
	0&=& \delta q-\kappa T_{0}\Big[-\frac{\nabla \delta T}{T_{0}}- \frac{\pd \delta v}{\pd t}-\beta_{1}\frac{\pd \delta q}{\pd t}+\alpha_{0}\nabla \delta \Pi+\alpha_{1}\vec{\nabla}.\delta\vec{\pi}\Big] \,. 
	\end{eqnarray}
	\end{subequations}
   We can further simplify the set of Eqs.~\eqref{eq58a}-\eqref{eq58f} by decomposing the fluid four velocity along the directions parallel and perpendicular to the 
	direction of wave vector, $\vec{k}$, and are termed as longitudinal component ($\delta \vec{v_{||}}$) and transverse component ($\delta\vec{v_{\perp}}$) 
respectively. The longitudinal and the transverse components of the fluid four velocity completely decouple the linearized 
hydrodynamic equations and we can get two sets linearly independent solutions for the two components. But the density perturbations 
appears only in the longitudinal component, which is our main focus here. Therefore, we consider the longitudinal component 
only for the present analysis. 
The hydrodynamic equations 
	can be solved for a given set of initial condition, $n(0),\, v_{||}(0),\, T(0),\, q(0),\, \Pi(0)$
	and  $\pi(0)$, by using the Fourier-Laplace transformation as:
	
	\beqa
	\delta Q( \vec{k}, \omega)= \int^{\infty}_{-\infty} d^3\vec{r} \int^{\infty}_{0}dt e^{-i(\vec{k}.\vec{r}-\omega t)} \delta Q(\vec{r}, t)\,.
	\label{eq18}
	\eeqa
	The $\delta P$ and $\delta s$ can be written in terms of the independent variables $n$ and $T$ as 
	follows by using the thermodynamic relations:
	\beqa
	\delta P&=&\Big(\frac{\pd P}{\pd n}\Big)_{T}\delta n+ \Big(\frac{\pd P}{\pd T}\Big)_{n}\delta T\,,\nn\\
	\delta s&=&\Big(\frac{\pd s}{\pd n}\Big)_{T}\delta n+ \Big(\frac{\pd s}{\pd T}\Big)_{n}\delta T\,.
	\label{eq19}
	\eeqa
	We use Eqs. \eqref{eq18} and \eqref{eq19}
	to write down the  longitudinal linearized hydrodynamic equation as: 
	\beqa
	\mathds{M}\delta Q(\vec{k}, \omega)=  \delta Q(\vec{k},0)\,,
	\label{eq20}
	\eeqa
	where, 

		\beqa
		\mathds{M} =
		\begin{bmatrix}
			-i\omega & ikn_{0} & 0 & 0 & 0  & 0   \\
			-i\omega \big\{n_{0}\big(\frac{\pd s}{\pd n}\big)_{T}- \frac{\mathcal{A}}{n_0\,T_0} \big\}   &    0 & -i\omega n_{0}\big(\frac{\pd s}{\pd T}\big)_{n} & 0  &  \frac{k}{T_0} &0   \\
			{ik}  \big(\frac{\pd P}{\pd n}\big)_{T}& -i\omega(h_0+\mathcal{A}) &  {ik}  \big(\frac{\pd P}{\pd T}\big)_{n}&  {ik}   & -{i\omega } &   i k \\
			0 & ik\zeta & 0 & (1-i\omega \beta_{0} \zeta) & - ik\alpha_{0} \zeta  & 0 \\
			
			0 & -i\frac{4}{3} \eta k & 0 & 0 &  \frac{4}{3}\alpha_{1}k \eta   & (1-2i\omega \beta_{2}\eta)\\
			0 & -i\omega \kappa T_{0} & ik\kappa &  -ik\alpha_{0} \kappa T_{0}& (1-i\omega \beta_{1} \kappa T_{0})  &- ik \alpha_{1} \kappa T_{0} \nn\\
		\end{bmatrix}\,,
		\\
		\label{eq21}
		\eeqa
and 
		\beqa
		\delta Q(\vec{k},\omega)=
		\begin{bmatrix}
			\delta n(\vec{k},\omega) \\
			\delta v_{||}(\vec{k},\omega)\\
			\delta T(\vec{k},\omega)\\
			\delta \Pi(\vec{k},\omega)\\
			\delta q_{||}(\vec{k},\omega)\\
			\delta \pi_{||}(\vec{k},\omega)
		\end{bmatrix}
		;\hspace{0.1cm}
		\delta Q(\vec{k},0)=
		\begin{bmatrix}
			\delta n(\vec{k},0) \\
			\Big\{n_{0}\Big(\frac{\pd s}{\pd n}\Big)_{T}- \frac{\mathcal{A}}{n_0\,T_0} \Big\}  \delta n  (  \vec{k},   0)+n_{0}\Big(\frac{\pd s}{\pd T}\Big)_{n}\delta T(\vec{k},0)\\ 
			(h_0+\mathcal{A})\delta v_{||}(\vec{k},0)+ \delta q_{||}(\vec{k},0)\\
			\beta_{0}\zeta \delta \Pi(\vec{k},0)\\
			2\beta_2 \eta \delta \pi_{||}(\vec{k},0)\\
			\kappa T_{0}\delta v_{||}(\vec{k},0)+\kappa T_{0}\beta_{1}\delta q_{||}(\vec{k},0)\\
			
		\end{bmatrix},
		\label{eq22}
		\eeqa 
		here, $\mathcal{A}=\,(B^2-MB)$. 

The solution for density fluctuation is obtained by solving the set of above algebraic equations which gives,

\beqa
\delta n(\vec{k},\omega)&=&\Big[ \mathds{M}^{-1}_{11}+\Big\{n_{0}\Big(\frac{\pd s}{\pd n}\Big)_{T}- \frac{\mathcal{A}}{n_0\,T_0} \Big\}\, \mathds{M}^{-1}_{12}\Big ]\, \delta n(\vec{k},0)
+\Big[n_{0}\Big(\frac{\pd s}{\pd T}\Big)_{n}\,\mathds{M}^{-1}_{12}\Big] \,{\delta T}(\vec{k},0) \nn\\
&&+\Big[(h_0+\mathcal{A})\,\mathds{M}^{-1}_{13} +\kappa T_{0} \mathds{M}^{-1}_{16} \Big]\,\delta v_{||}(\vec{k},0)
 - \Big[\mathds{M}^{-1}_{13}+\kappa T_{0}\beta_{1}\,\mathds{M}^{-1}_{16}\Big]\,\delta q_{||}(\vec{k},0) \nn\\
&&+\Big[\beta_{0}\zeta\,\mathds{M}^{-1}_{14}\Big] \delta \Pi (\vec{k},0)
+  \Big[ 2\beta_{2}\eta\,\mathds{M}^{-1}_{15} \Big]\,\delta \pi_{||}(\vec{k},0)\,.
\label{eq23}
\eeqa
The appearance of ${\mathcal{A}}$ in the density fluctuation  indicates the presence 
of magnetic field in the system. The expression of density fluctuation is a function of the other fluctuating hydrodynamic variables such as $\delta T,\, \delta v_{||},\,\delta q_{||},\,\delta \Pi$, and $\delta \pi_{||}$. The independence among hydrodynamic variables does not necessarily imply the absence of correlations~\cite{Jeon:2015dfa}. However, in this work, we have specifically assumed that there are no correlations between density fluctuations and other fluctuating hydrodynamic variables~\cite{Minami:2009hn,Hasanujjaman:2020zex}. Therefore, the correlation between two independent thermodynamic variables, say, $Q_i$ and $Q_j$  vanishes {\it{i.e.}},
\beqa
\Big< \delta Q_{i}(\vec{k},\omega)\delta Q_{j}(\vec{k},0)\Big>=0, \,\,\,\, i\neq j
\label{eq25}
\eeqa

The required correlator, $\mathcal{S^\prime}_{nn}(\vec{k},\omega)$ is obtained as:
\beqa
\mathcal{S^\prime}_{nn}(\vec{k},\omega)&=&\Big< \delta n(\vec{k},\omega)\delta n(\vec{k},0)\Big>\nn\\
&=&\Big[ \mathds{M}^{-1}_{11}+\Big\{n_{0}\Big(\frac{\pd s}{\pd n}\Big)_{T}- \frac{\mathcal{A}}{n_0\,T_0} \Big\}\, \mathds{M}^{-1}_{12}\Big ]\, \Big< \delta n(\vec{k},0)\delta n(\vec{k},0)\Big>\,.
\eeqa
Finally, the $\Snn$ is defined as: 
\begin{equation}
\mathcal{S}_{nn}(\vec{k},\omega)=\frac{\mathcal{S^\prime}_{nn}(\vec{k},\omega)}
   {\Big< \delta n(\vec{k},0)\delta n(\vec{k},0)\Big>}\,.
\label{eq26}
\end{equation}
The $\Snn$ contains the transport coefficients such as $\eta,\,\zeta,\,\kappa$ and other thermodynamic response functions, which will be used to study the behaviour of the structure factor.
It is well known that the relativistic Navier-Stokes (NS) hydrodynamic equation can be obtained 
by setting the various coupling ($\alpha_0$, $\alpha_1$) and relaxation 
($\beta_0$, $\beta_1$, $\beta_2$) 
coefficients to zero. Therefore, the $\Snn$ for NS hydrodynamics can be obtained 
under the similar limits as:
\beqa
\Snn & = & \frac{\mathcal{A} k^2 \chi  \omega-\omega^{2}h_{0}T)
(\frac{\pd s}{\pd T})_n -k^{2}\chi \omega h_{0}+k^{2}(\mathcal{A}-\omega \chi T)(\frac{\pd P}{\pd T})_n}{(\mathcal{A}-2) k^2 \chi\omega^2+k^2\chi \omega T(4 \eta/3
- \zeta )+(\frac{\pd s}{\pd T})_n \{2k^2 \omega^2 (4 \eta/3 -\zeta )-2T^{2} \chi  \omega^4\}}\,.
\eeqa
The appearance of higher order derivatives in MIS hydrodynamics makes the dispersion relation
a quintic equation in $\omega$ and the corresponding dispersion equation in NS hydrodynamics
is a cubic equation in $\omega$. 

The possibility of the existence of the critical end point (CEP) 
in the QCD phase diagram ~\cite{Fodor:2001pe,Fodor:2004nz} is considered as one of the most
interesting development in the field of relativistic 
heavy ion research. The location of the CEP in $T-\mu$ plane is not known
from first principle. The model dependent predictions  
vary widely as indicated in Ref.~\cite{DeWolfe:2010he}. 
In the present work we chose ($T_{c},\,\mu_{c})= (154,\,367)$MeV to study
the effects of the CEP on the spectral function in the
presence of external magnetic field, $\vec{B}$. The effects of
CEP are included in the calculation of spectral function
through the equation of state~\cite{Nonaka:2004pg,Parotto:2018pwx} 
and scaling behaviour
of the transport coefficients~\cite{Guida:1996ep,Rajagopal:1992qz,Kapusta:2012zb}. The details of
this part of the calculations are given in Refs. ~\cite{Hasanujjaman:2020zex,Sarwar:2022iem}.
Therefore, we refer to these references for details to avoid repetition.

\section{Results} 
Before we present the results it is important to mention the following points.
In the context nuclear collisions at relativistic energies one needs to 
solve the relativistic viscous second order hydrodynamical equations in the presence of the CEP
and ultra-high magnetic field. Development of such a numerical code is highly time consuming which will be considered in future publications.
In the present work we focus on the study of the dynamical density correlation in a non-expanding QGP in the presence 
of an ultra-high magnetic field with and without the inclusion of the CEP. This study is important to 
check whether the magnetic field 
alters the nature of the correlation  near the CEP.  We find that the nature of the correlation near the CEP marked by the
absence of the Brillouin peaks remains unaltered in presence of the magnetic field. This is one of the important observation
of the present work as shown below.

It may also be mentioned here that the matter produced after nuclear collisions at LHC and top RHIC energies will be gluon dominated initially, however, 
quarks-antiquarks pairs will be produced dynamically due to the interactions among the gluons. Therefore, if the magnetic field ($B$)
does not decay (which depends on the electrical conductivity of the QGP) completely before the generation of quark-antiquark pairs 
then the effects of $B$ will be realized. Moreover, at lower RHIC energies (Beam Energy Scan Program) the quarks and
antiquarks will be present from the very early stage and will get affected by the magnetic field.  
\label{sec:sec4}
\begin{figure}
	\centering
	\includegraphics[width=0.66 \textwidth]{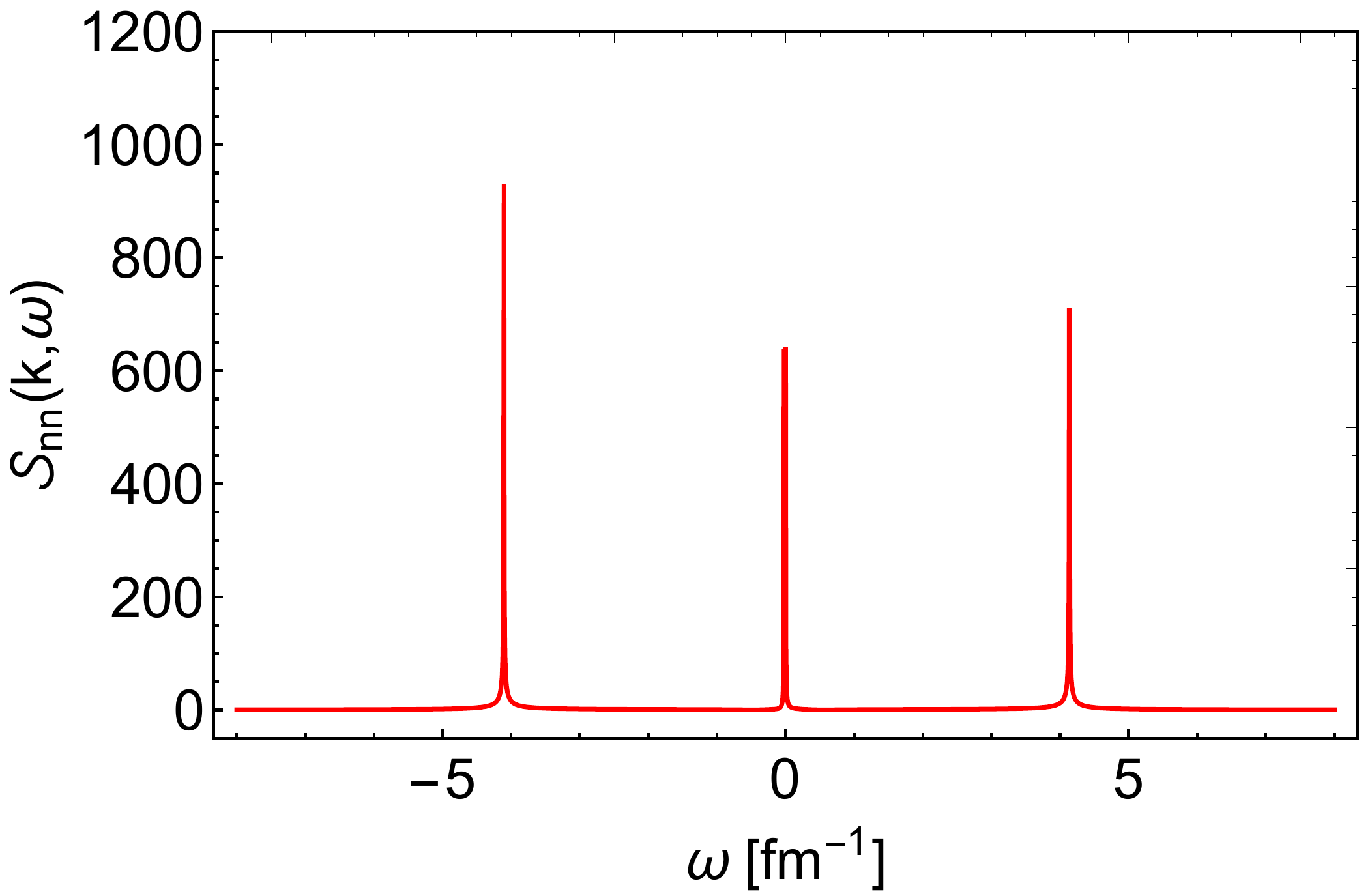}
	\caption{{\it{(Color online) The variation of the $\Snn$ with $\omega$ is shown with magnetic field strength, $eB=m_{\pi}^{2}$ for $k=0.1$ fm$^{-1}$ at $r=0.2$. It shows one R-peak and two B-peaks, symmetrically located about $\omega=0$.}}}
	\label{fig1}
\end{figure} 
\begin{figure}
	\centering
	\includegraphics[width=0.6\textwidth]{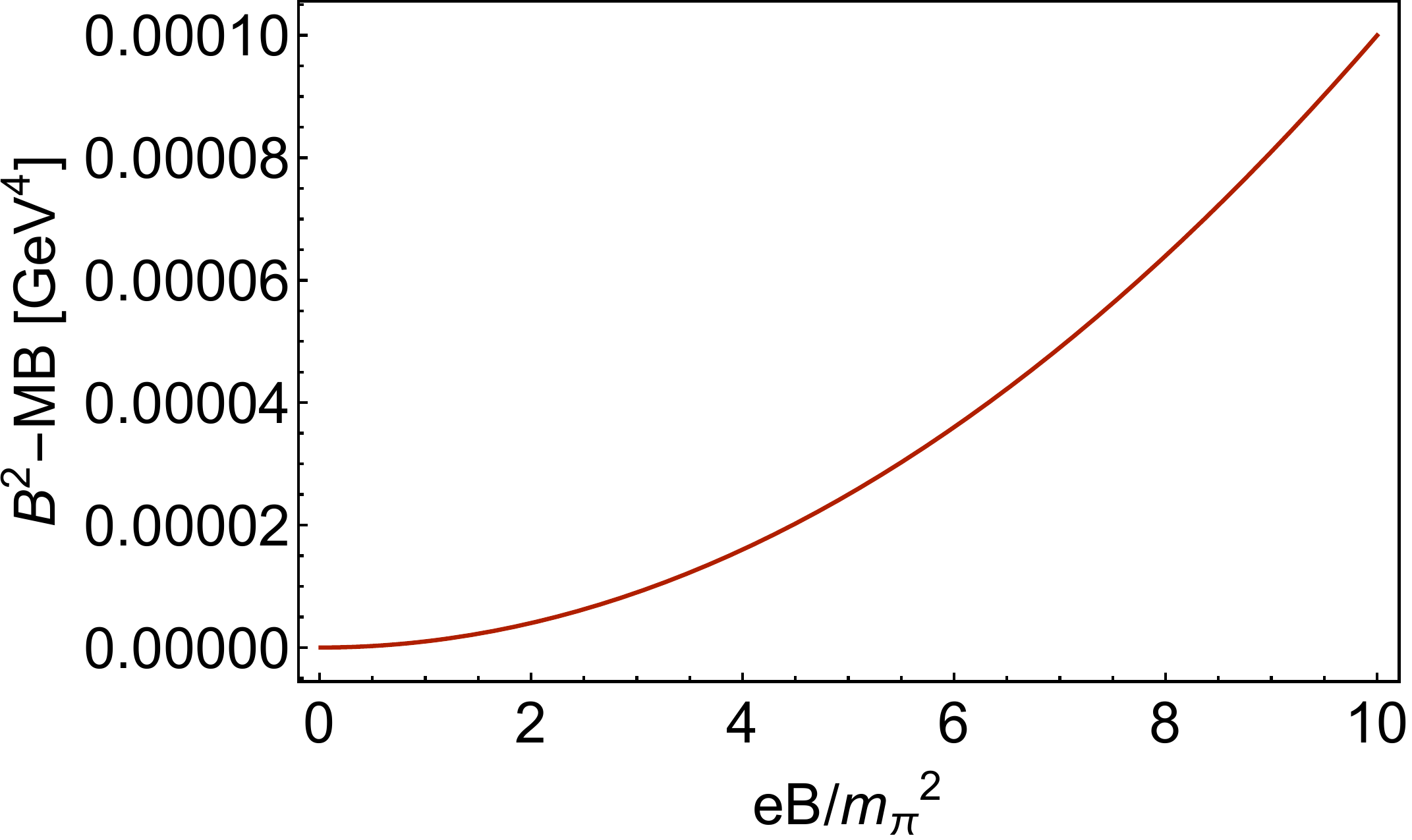}
	\caption{{\it{(Color online) The factor $B^{2}-MB$ is plotted with the variation of magnetic field.}}}
	\label{fig2}
\end{figure}

We aim to study the nature of the dynamic structure factor [$\Snn$] in the presence of a 
static magnetic field, $B$ here. The $\Snn$ in the absence of $B$ has been studied in earlier works~\cite{Hasanujjaman:2020zex,Sarwar:2022iem}, where it has been shown that the $\Snn$ admits three identifiable peaks. The central Rayleigh peak (R-peak) is positioned at angular frequency $\omega=0$, and the other two peaks, the Brillouin peaks (B-peaks) are situated on both sides of the R-peak with even magnitudes. The R-peak and the B-peaks are originated from the entropy (thermal) fluctuation at constant pressure and pressure fluctuation at constant entropy respectively. The position of the B-peaks enables us to evaluate the speed of sound. Also, the width and the integrated intensities of those peaks are associated with various thermodynamic quantities such as the isothermal compressibilities and specific heats of the system~\cite{Stanley_book}. 

Fig.~\ref{fig1} shows the $\Snn$ for $eB=m_{\pi}^{2}$ at $r=(T-T_{c})/T_{c}=0.2$, i.e. when the system is away from the CEP. The transport coefficients are taken as $\eta/s=\zeta/s=\chi\,T/s=1/4\pi$. We see three different peaks which are recognized as the R-peak (central), and the other two peaks as the B-peaks. The B-peaks are positioned symmetrically about $\omega=0$, but their heights are not identical. The 
unequal height of the B-peaks may occur due to the local inhomogeneity present in the system. The B-peaks arise from propagating sound modes associated with pressure fluctuations at constant entropy. In condensed matter physics, the asymmetry of the B-peaks is identified from the fact that two sound modes with different $\omega$ values, $-c_sk$ and $+c_sk$ originate from different temperature zones~\cite{Rayleigh_Benard,Zarate}. Furthermore, we see that the widths of the peaks are very narrow, implying the slow relaxation of the thermal as well as the pressure fluctuations, which keeps the system to linger at out-of-equilibrium state for a longer time.\begin{figure}
	\centering
	\includegraphics[width=0.6 \textwidth]{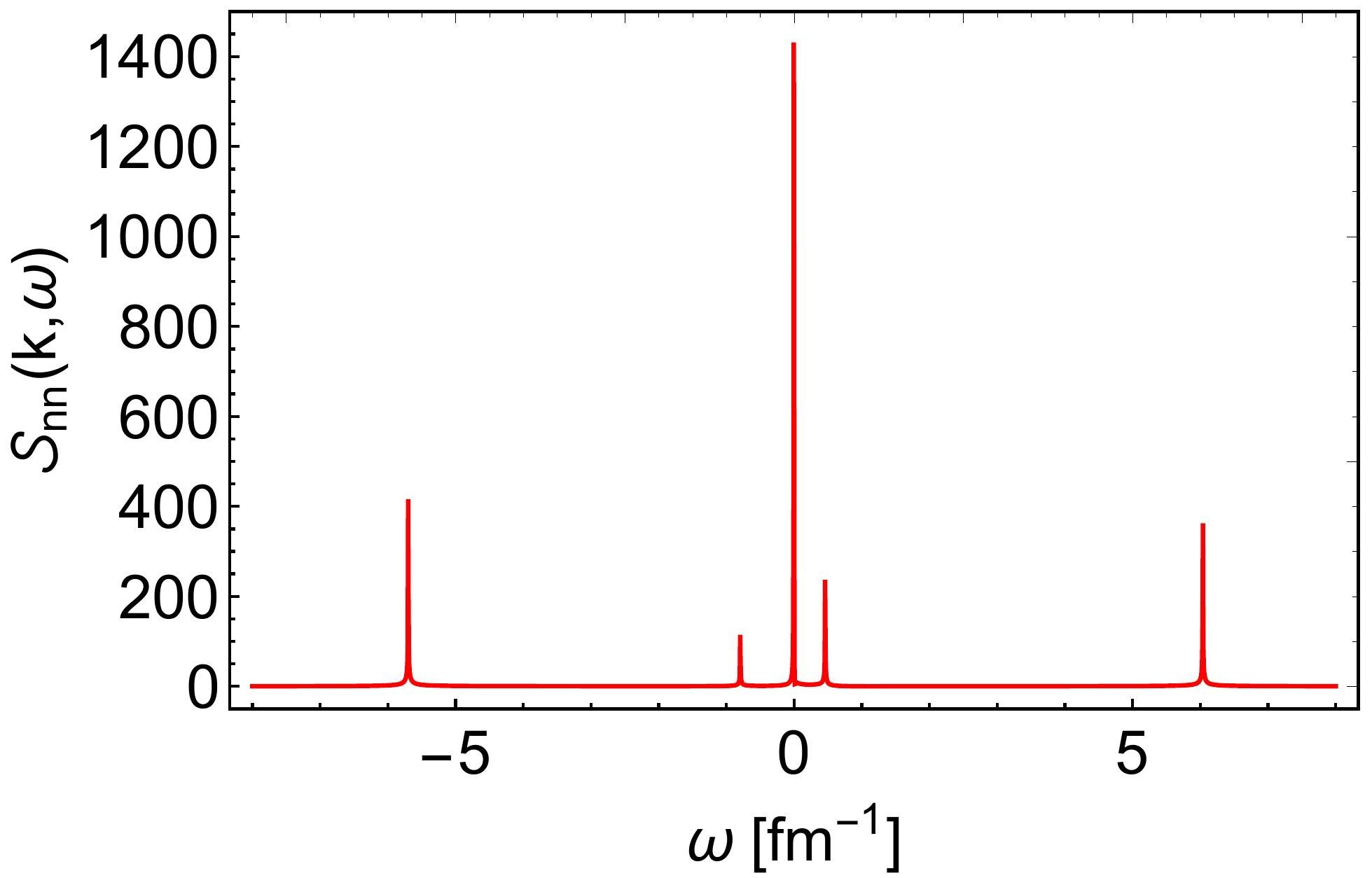}
	 	\caption{{\it{(Color online) The variation of the $\Snn$ with $\omega$ is shown in the presence of a magnetic field ($B=3\,m^{2}_{\pi}$) for $k=0.1$ fm$^{-1}$ at $r=0.2$. It shows one R-peak, two away side B-peaks asymmetrically located about $\omega=0$, and two near side B-peaks, originated from the coupling of the magnetic field with the other thermodynamic fields.}}}
	\label{fig3}
\end{figure} 

The introduction of $B$ results in splitting the pressure into transverse, $P_\perp=P-MB$ and longitudinal, $P_L=P$ components. 
The expression of the magnetization is adapted from Ref.~\cite{Huang:2009ue}. 
The value of $\mathcal{A}$ determines the effects of the magnetic field on the spectral function as
evident from the expression of $\delta n$ in Eq.~\eqref{eq23}. Therefore,
we show the variation of $\mathcal{A}=B^2-MB$ with $B$  in Fig.~\ref{fig2}. The small value of $\mathcal{A}$ 
for low $B$ indicates that the effects of $B$ on $\Snn$ will be significant only beyond a certain value of $B$.

The $\Snn$ in the presence of the magnetic field with $eB=3\,m_{\pi}^{2}$ is shown in Fig.~\ref{fig3}, where we see five distinct peaks. The peak found at $\omega=0$ is identified as the R-peak. The two distant peaks about the R-peak are recognized as the B-peaks are originated from the thermodynamic pressure fluctuation at constant entropy (longitudinal component). The magnitudes of the B-peaks are not even as well as their position about $\omega=0$ is not symmetric. The positional asymmetry could be realized due to the presence of a unidirectional magnetic field. The other two nearby peaks around the R-peak are also found to be the B-peaks arises from the 
pressure fluctuation in the transverse direction. These B-peaks only appear in the presence of the 
magnetic field beyond a certain strength. The threshold value for the emergence of the B-peaks (near) is found to be 
at $B_{th}=2.32\,m_{\pi}^{2}$. The transverse pressure ($=P-MB$) in Eq.~\eqref{eq39},
containing the $B$ field is solely responsible for the appearance of the near side B-peaks.
\begin{figure}
	\centering
	\includegraphics[width=0.6
	\textwidth]{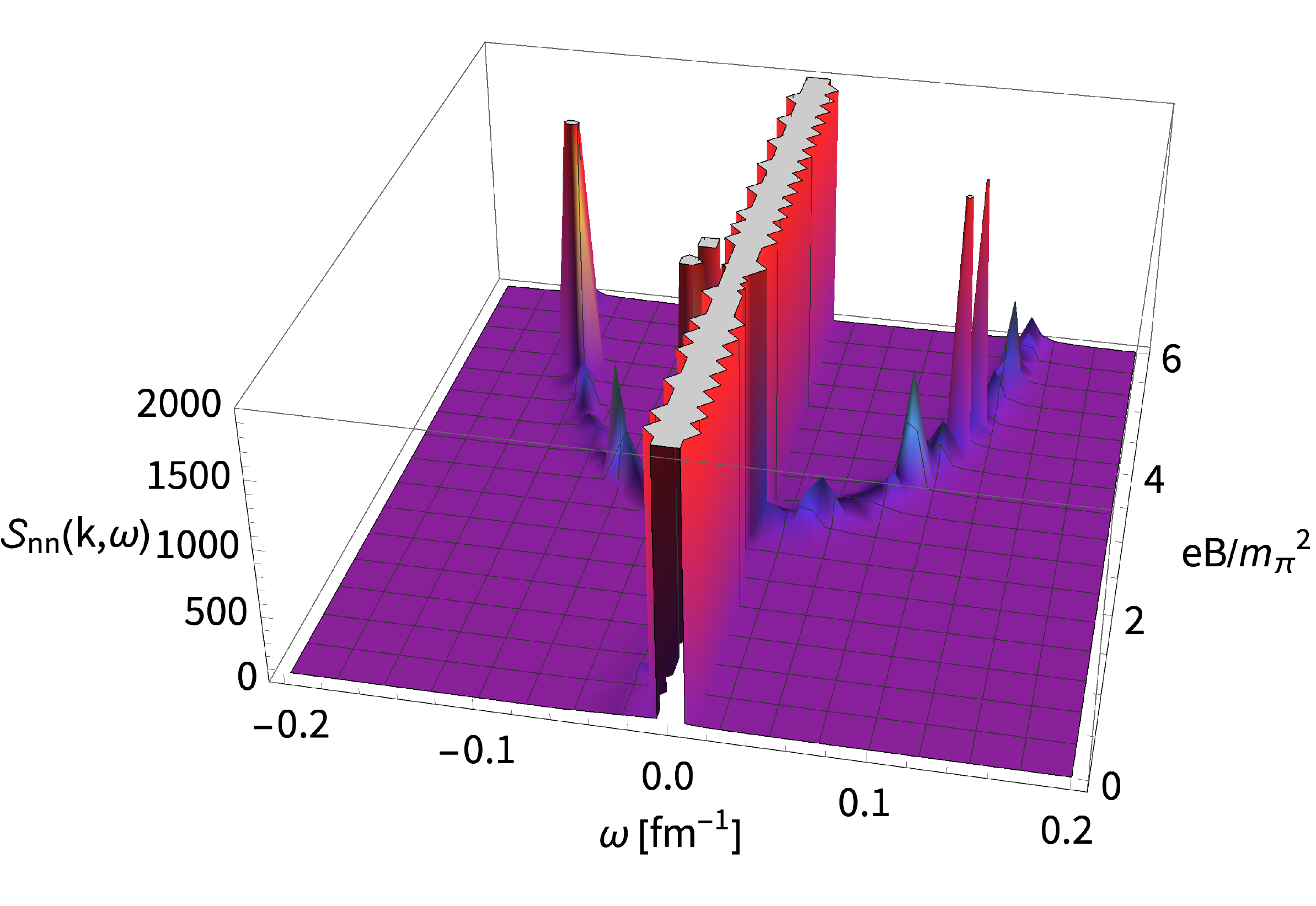}
	\caption{{\it{(Color online) A 3D plot is shown to understand the role of the magnetic field in emergence of the near side B-peaks. The peaks start appearing after a threshold value of the magnetic field, $B_{th}$.}}}
	\label{fig4}
\end{figure}

The importance of the magnetic field in the appearance of those extra B-peaks can be  
understood from Fig.~\ref{fig4}, where the $\Snn$ is plotted 
as a function of $B$. The range of $\omega$ is kept small to focus on the near side B-peaks only. 
For the smaller values of $B$, no extra peak is found, and they start appearing after a threshold value of $B_{th}$.

\begin{figure}
	\centering
	\includegraphics[width=0.6 \textwidth]{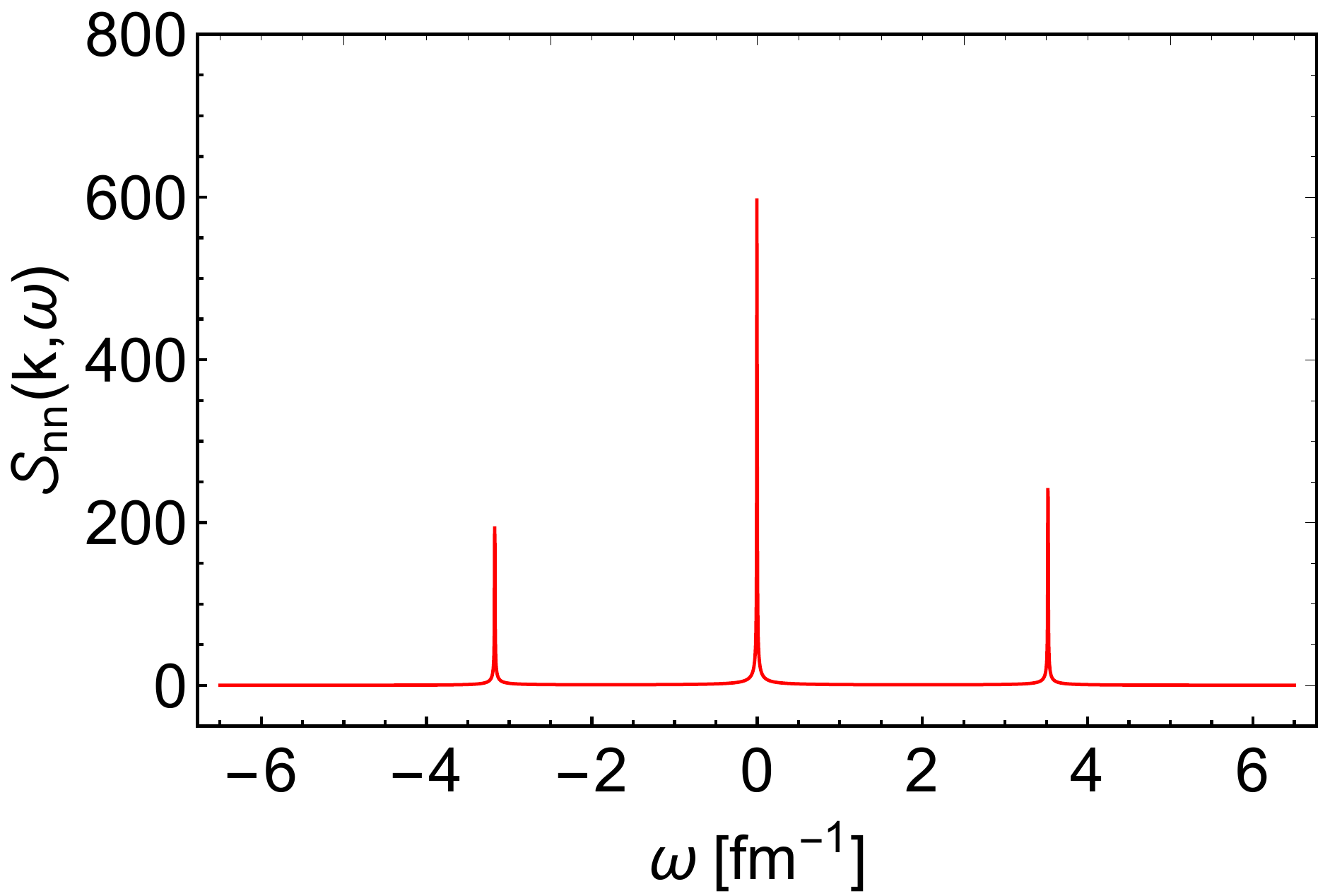}
	 	\caption{{\it{(Color online) The variation of the $\Snn$ with $\omega$ is shown in the presence of a magnetic field in Navier-Stokes theory. It shows one R-peak and two away side B-peaks asymmetrically located about $\omega=0$.}}}
	\label{fig5}
\end{figure}

Fig.~\ref{fig5} shows the $\Snn$ in the presence of the magnetic field ($eB=3\,m_{\pi}^{2}$) in NS theory, which is recovered 
from the MIS model by putting the relaxation and coupling coefficients to zero. We see only three peaks, where the extra peaks due to the presence of the magnetic field are not appearing, but the presence of the magnetic field is evident from the magnitude as well as positional asymmetry of the B-peaks. Therefore, we argue that the extra near-side B-peaks are generated due to the coupling of hydrodynamic fields with the magnetic field when we consider second-order hydrodynamics. 
This is understood because the dispersion relations of MIS and NS hydrodynamics are
quintic and cubic equations respectively. 

\begin{figure}
	\centering
	\includegraphics[width=0.45 \textwidth]{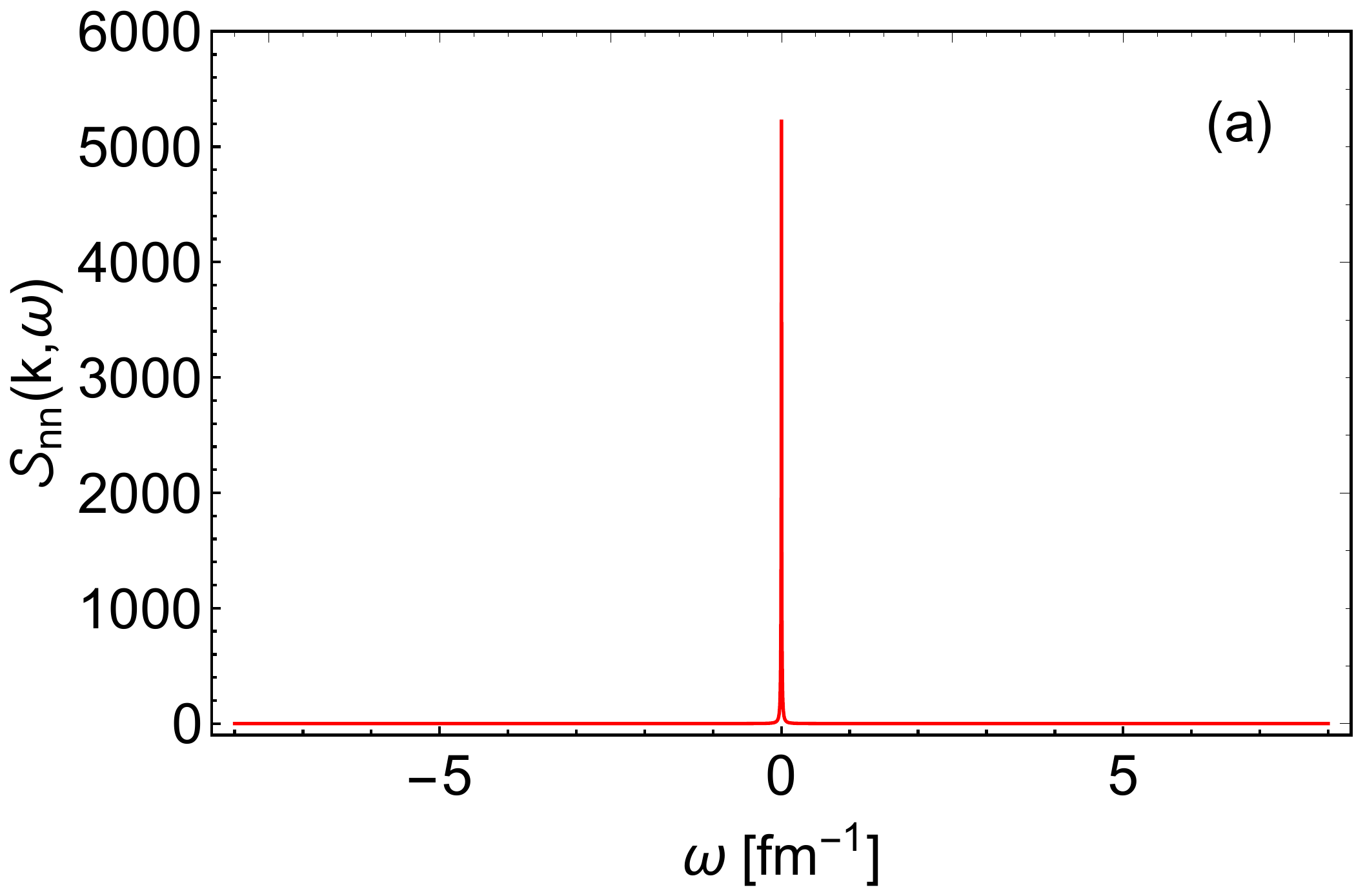}
	\includegraphics[width=0.45 \textwidth]{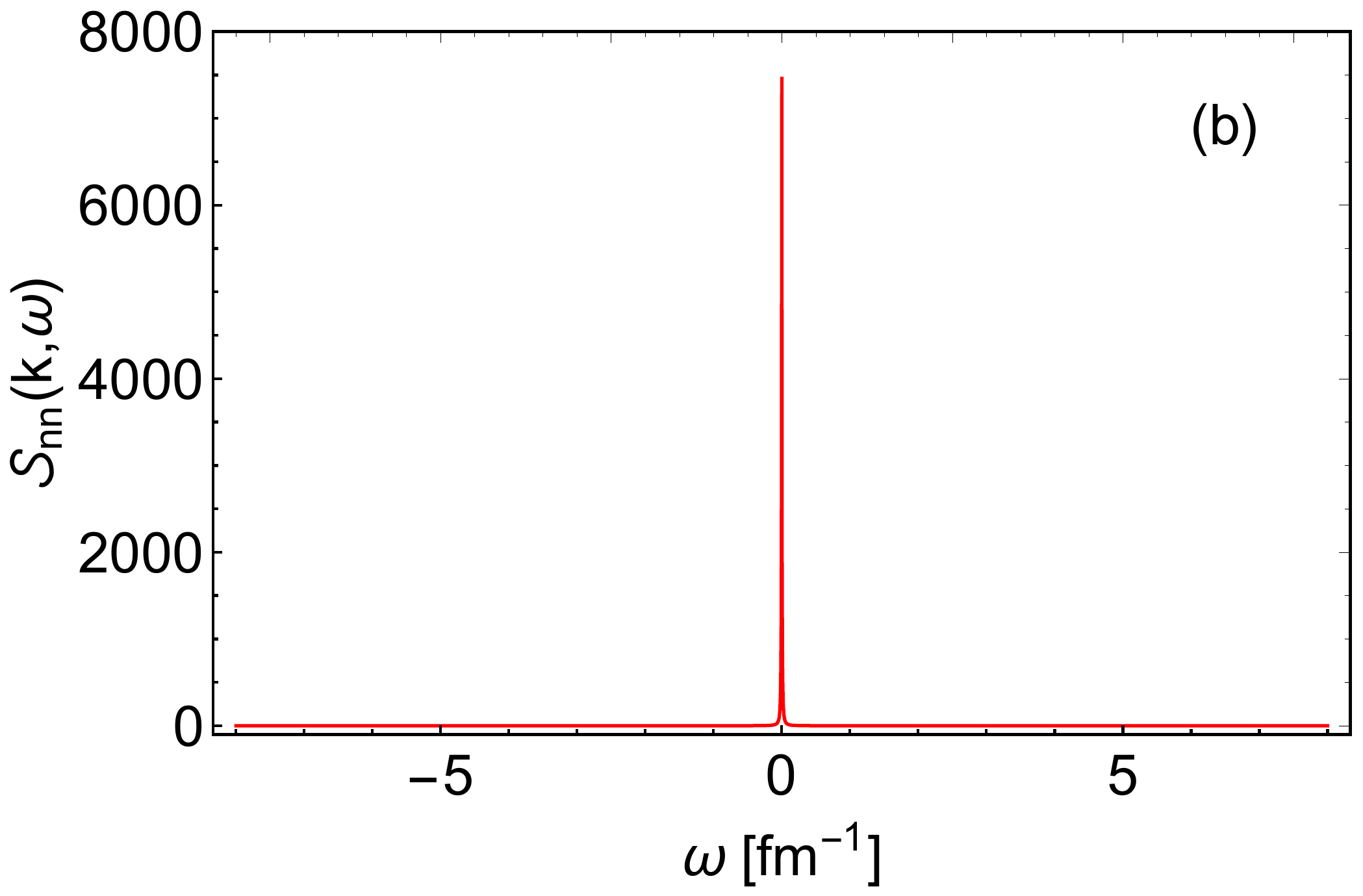}
	 	\caption{{\it{(Color online) The variation of the $\Snn$ with $\omega$ near the critical end point for $k=0.1$ fm$^{-1}$ at $r=0.01$ is shown in the presence of a magnetic field in MIS theory for a) magnetic field strength $B=m^{2}_{\pi}<B_{th}$, and b) magnetic filed strength $B=3\,m^{2}_{\pi}\ge B_{th}$.}}}
	\label{fig6}
\end{figure}

The behaviour of the structure factor near the CEP ($r=0.01$) is shown in Fig.~\ref{fig6}. The transport coefficients and the thermodynamics response functions that appear in Eq.\eqref{eq26} are taken from the scaling laws~\cite{Guida:1996ep,Rajagopal:1992qz,Kapusta:2012zb,Hasanujjaman:2020zex,Sarwar:2022iem}.
As the near side B-peaks appear only after a threshold intensity of the magnetic field, $B_{th}$, we have plotted the $\Snn$ 
for two different values of magnetic field: a) $B=m^{2}_{\pi}$, and b) $B=3\,m^{2}_{\pi}$. In both cases, we do not observe any B-peaks but only the R-peak with 
a larger magnitude. This happens due to the absorption of sound near the CEP and hence 
B-peaks vanish. 
Comparing both the cases, we also observe that the magnitude of the R-peak in Fig.~\ref{fig6}(b) is 
larger compared to the R-peak in Fig.~\ref{fig6}(a). 
The most important observation here is that the B-peaks disappear near the CEP irrespective of the value of the external 
fields (zero or non-zero).  

\section{Summary and conclusion}
\label{sec:conclusion}
The dynamic structure factor, $\Snn$, is evaluated theoretically from the correlation of the density fluctuation, when the system with
a CEP is subjected to an external static magnetic field. Without any magnetic field, the $\Snn$ admits a Rayleigh peak and 
two Brillouin peaks. But in presence of the magnetic field, we find two extra peaks appearing closer to the  R-peak which 
are identified also as the B-peaks due to the adiabatic 
transverse pressure fluctuation. The asymmetry in magnitudes of the B-peaks (away side) are realized due to the local 
inhomogeneity of the system, whereas, the positional asymmetry appears due to the presence of the magnetic field. 
The extra B-peaks at lower $\omega$ exclusively appear beyond a threshold value of the magnetic field. 
The magnetic field splits the pressure into transverse, $P_\perp=P-MB$ and longitudinal, $P_L = P$ components. 
The away side and the near side B-peaks are caused by the pressure fluctuations in longitudinal and transverse direction respectively. 
To understand the role of the relaxation and the coupling coefficients, we have 
evaluated $\Snn$ for NS theory and find two away side B-peaks, positioned asymmetrically with uneven 
heights. Therefore, we argue that the near side B-peaks appear due to the coupling of the magnetic field with the hydrodynamic 
fields in the second-order hydrodynamic theory. 
It will be interesting to investigate the role of this asymmetry in a realistic scenario with time varying electromagnetic
field in an expanding QGP with CEP, which is beyond the scope of the present theoretical case study performed in an idealistic scenario
of non-expanding QGP in a static magnetic field.

\section{Acknowledgement}
\label{sec8}
MR, MH would like to thank Department of Higher Education, Govt. of West Bengal, India for the support. 
\bibliography{spec_B.bib} 
\end{document}